\documentclass[rnote]{aa}

\usepackage{graphicx,color}
\usepackage{natbib}
\usepackage{txfonts}
\usepackage{amsmath,amssymb}
\bibpunct{(}{)}{;}{a}{}{,}

\begin{document}

\title{Silicate condensation in Mira variables}


\author{
 Hans-Peter Gail\inst{1}
 \and Michael Scholz\inst{1,2}
 \and Annemarie Pucci\inst{3}
}

\institute{
Universit\"at Heidelberg, Zentrum f\"ur Astronomie, 
           Institut f\"ur Theoretische Astrophysik,
           Albert-Ueberle-Str. 2,
           69120 Heidelberg, Germany 
\and
           University of Sydney, Sydney Institute for Astronomy,
           Sydney NSW 2006, Australia
\and
Universit\"at Heidelberg, Kirchhoff-Institut f\"ur Physik,
           Im Neuenheimer Feld 227,
           69120 Heidelberg, Germany 
  }

\offprints{\tt gail@uni-heidelberg.de}

\date{Received date ; accepted date}

\abstract{The formation of dust in the winds of cool and highly evolved stars and the rate of injection of dust into the ISM is not yet completely understood, despite the importance of the process for the evolution of stars and galaxies. This holds in particular for the oxygen-rich stars where it is still not known which process is responsible for the formation of the necessary seed particles of their silicate dust. 
}
{We study whether the condensation of silicate dust in Mira envelopes could be caused by cluster formation by the abundant SiO molecules.
}
{For a simplified model of the pulsational motions of matter in the the outer layers of a Mira variable which is guided by a numerical model for Mira pulsations, the equations of dust nucleation and growth are solved in the co-moving frame of a fixed mass element. It is assumed that seed particles form by clustering of SiO.  The calculation of the nucleation rate is based on the experimental findings of Nuth \& Donn (1982). The quantity of dust formed is calculated by a moment method and the calculation of radiation pressure on the dusty gas is based on a dirty silicate model. 
}
{Dust nucleation occurs in the model at the upper culmination of the trajectory of a gas parcel where it stays for a considerable time at low temperatures. Subsequent dust growth occurs during the descending part of the motion and continues after the next shock reversed motion. It is found that sufficient dust forms that radiation pressure exceeds gravitational pull of the stars such that the mass element is finally driven out of the star. 
}
{Nucleation of dust particles by clustering of the abundant SiO molecules could be the mechanism that triggers silicate dust formation in Miras.
}


\maketitle

\titlerunning{Dust condensation in Miras}

\section{Introduction}

Outflow of matter from highly evolved stars is generally observed to be accompanied by copious dust formation. Since the gas streaming out of a stellar atmosphere is free from condensation seeds, the first step in the dust condensation process has to be the formation of seed particles, followed by precipitation of condensable material on the freshly created surfaces. What one directly sees in observations is the outcome of the second step, the growth of dust grains. In case of late-type M stars one observes that the dominating dust components formed in the outflow are magnesium-iron-silicates \citep[see][ for a review on observations]{Mol10}. Laboratory investigations of presolar silicate dust grains from AGB stars have shown that usually Mg-Fe-silicates are formed \citep{Vol09a,Ngu10,Bos10} that have widely varying iron content, where iron-poor grains are rare.

The first step, the nucleation process, is not observable (at least for the moment). Because SiO is one of the most abundant of all the gas-phase species of refractory elements that could be involved in this process, it has been often discussed whether cluster formation by this species is the starter reaction for the sequence of processes that ultimately leads to the condensation of silicate dust particles \citep{Don81, Nut82, Nut83, Gai86, Gai98, Gai98b, Ali05, Nut06, Reb06, Paq11, Gai13}, though other possibilities have recently also be discussed \citep{ Gou12, Pla13, Bro14,Gob16}. 

In \citet{Gai13} we have studied nucleation of SiO as the primary dust formation stage in oxygen-rich environments on the basis of our recent measurements of the vapour pressure of solid SiO \citep{Wet12}. We derived an empirical nucleation rate for SiO from a re-evaluation of the experimental data for SiO nucleation of \citet{Nut82}. By calculation of stationary wind models with nucleation and dust growth we found that SiO nucleation as the trigger for the condensation of silicate dust in M stars is not as unlikely as previously thought. The previous calculations \citep{Gai86} resulted in far too low condensation temperature because they were based on a vapour pressure measurement for SiO \citep{Sch60} that yielded unrealistically high values. With revised data on SiO vapour pressure the onset of formation of dust occurs already at temperatures close to dust temperatures of 800 K \dots\ 950 K at the inner edge of dust shells as derived by fitting radiative transfer models to observed infrared spectra from dusty stars \citep[e.g.][]{Gro09}. These results suggest that nucleation by SiO may in fact provide the necessary seeds for silicate dust growth. A similar conclusion was arrived at in a different way by \citet{Nut06}.

While stationary wind models may be fair approximations for the outflows from supergiants which do not show strong variations and no shock structure in their upper atmospheres \citep{Arr15}, the atmospheres of Mira variables are characterized by large pulsational excursions of the matter in the outer atmospheres and envelopes during a pulsation cycle, and by series of shocks propagating outward. Miras represent the largest group of dust enshrouded oxygen-rich stars, thus it is important to know their dust formation mechanisms. Therefore we attempt to model dust condensation in such objects based on the assumption that cluster formation by SiO is the trigger for silicate dust formation. As a first step we study a simple ballistic model for the motion of the gas during a pulsation cycle and study dust formation by following a fixed gas parcel. The ballistic model is guided by the Mira pulsation models described in \citet{Ire06,Ire08,Ire11}. A consistent implementation of dust formation in the pulsation code would be desirable, but this is presently not available.    

We will show that also for pulsating stars the assumption of seed particle formation by SiO nucleation seems to yield results that are in accord with what one observes for dust forming Miras. This suggests that SiO nucleation provides the seed particles for silicate dust formation also in that case. 

\section{The model}

\subsection{Mira model}
\label{SectMiraMod}

\begin{figure}
\resizebox{\hsize}{!}{\includegraphics[angle=-90]{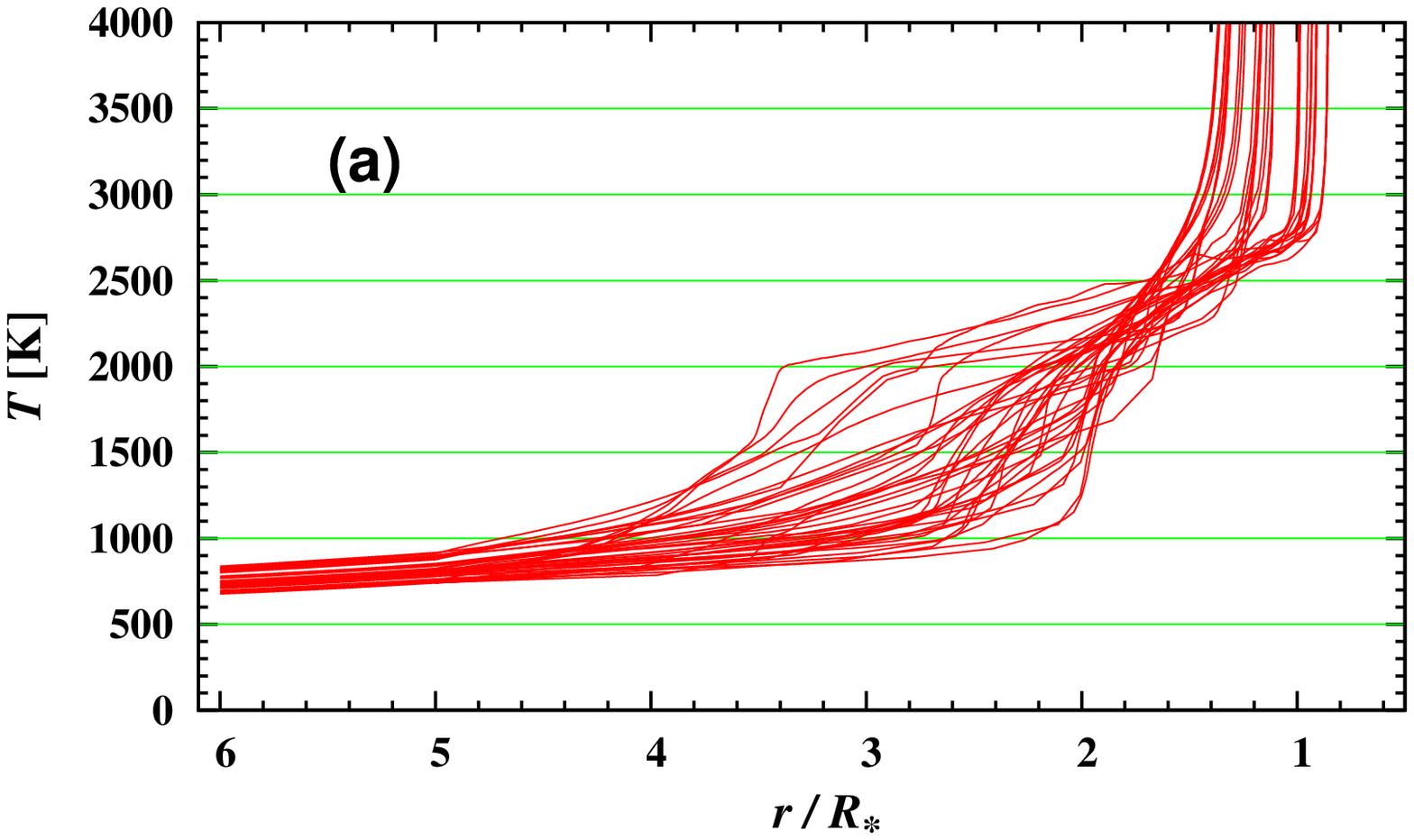}}

\resizebox{\hsize}{!}{\includegraphics[angle=-90]{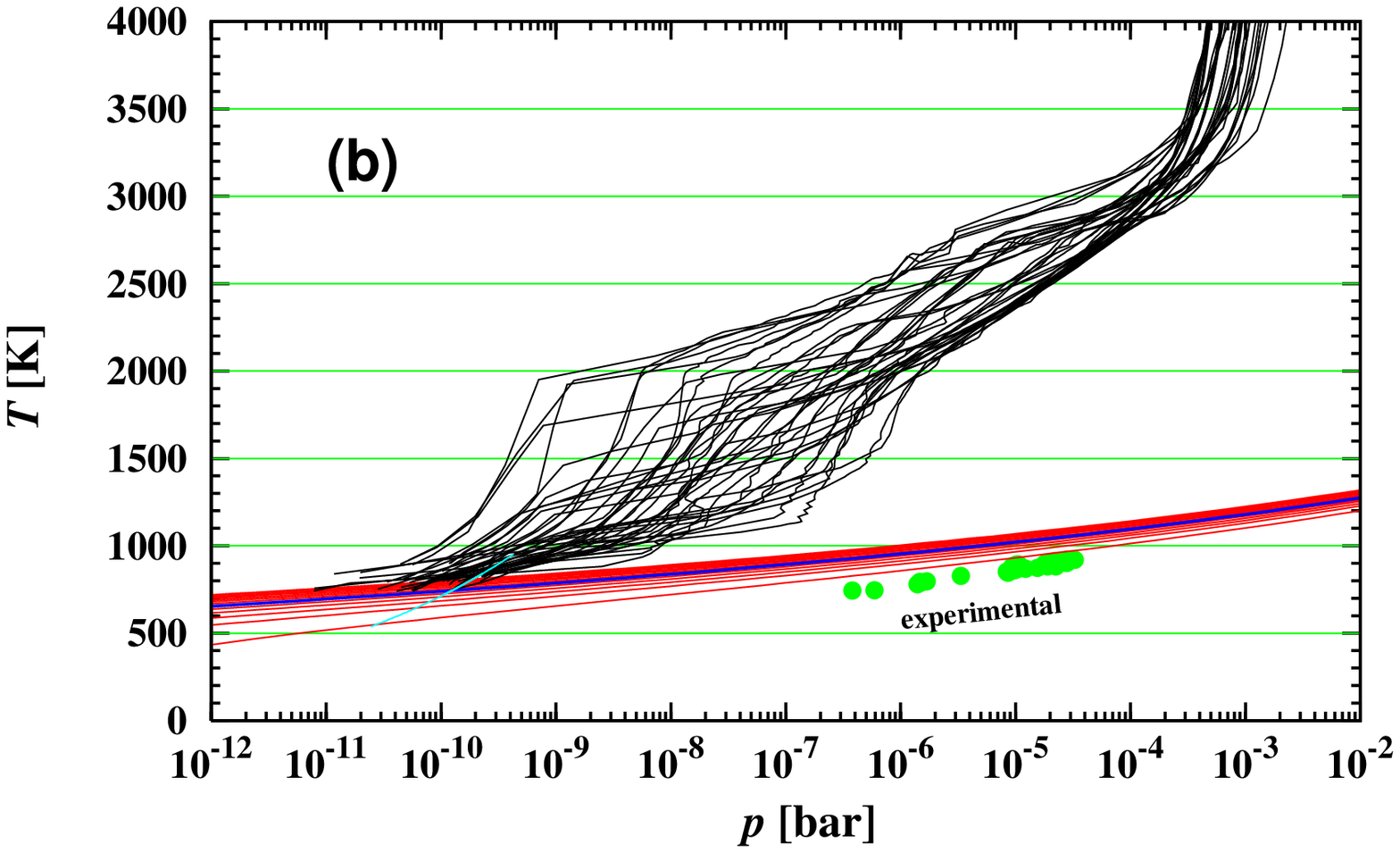}}

\caption{(a) Radial variation of temperature for a number of pulsation phases covering the phase range from $\Phi=-0.20$ to $\Phi=3.61$ (extrapolated between 5 an $6\,R_*$). (b) Run of pressure and temperature for the pulsation model at different phases of the pulsation cycle between $\Phi=-0.2$ and $\phi=3.8$ (solid black lines). The nearly horizontal red lines correspond to lines of constant normalized nucleation rate with values between $10^{-34}$ and $10^{-12}$ particles per second and per H nucleus in steps of $10^ {-2}$. The blue line corresponds to the onset of avalanche nucleation with a rate of $10^{-20}$ s$^{-1}$ and per H nucleus. The experimental data of \citet{Nut82} are indicated by green circles. The green line corresponds to the upper green line in Fig.~\ref{FigTraj}.}

\label{FigNukPMod}

\end{figure}

We assume for this study for simplicity spherical symmetry of the star and its outflow and use a typical atmospheric model of the CODEX series published by \citet{Ire08, Ire11}. The  CODEX models cover several
cycles of Mira variables, based on a self-excited pulsation model, for
four different non-pulsating "parent stars" (mass $M$, luminosity $L_\mathrm{p}$,
radius $R_\mathrm{p}$). Solar element abundances were assumed \citep[cf.][ for effects of moderate abundances changes]{Sch14}. The non-grey atmospheric stratification was computed up to 5 $R_\mathrm{p}$ above which a variation $T(r) \propto r^{-1/2}$ with radial distance $r$ may be assumed. We choose for the present study the four-cycle tight phase coverage (39 phases from -0.20 to 3.61) of the model Mira o54 the parent star of which has parameters close to the Mira prototype $o$~Cet ($M = 1.1\,\rm M_{\odot}$, $L_\mathrm{p}=5400\,\rm L_{\odot}$, $R_\mathrm{p}=216\,\rm R_{\odot}$). The luminosities $L$, the
$\tau_{\rm_{Ross}}=1$ Rosseland stellar radii $R$ and the effective
temperatures $T_{\rm eff} \propto (L/R^{2})^{1/4}$ of these
39 atmospheric models, as well as the positions of shock fronts at each pulsation 
phase, are given in Table 4 of Ireland et al. (2011). Figure \ref{FigNukPMod} shows the temperature stratifications of these models.

Since test calculations show that, for following the formation and final
outflow of typical dust particles, a time interval of about 5 pulsation
periods is needed, we extended the phase coverage (-0.20 to 3.61) of the
o54 model by assuming that model stratifications at phases 3.80 to 5.00
may be approximated by stratifications at phases -0.20 to 1.00. In fact,
inspection of shock fronts and stratifications shows that, by using this
phase extrapolation, one obtains a smooth and very reasonable transition
from phase 3.61 to phase 3.80 which should be perfectly sufficient for
studying the typical behaviour of a dust particle over 5 pulsation periods.

Figure \ref{FigNukPMod} shows, for phases -0.20 to 3.60, the temperature as function of position $R/R_\mathrm{p}$ and of pressure $p$. We note that the typical
decrease of $T(r)$ is strongly related to the position of shock fronts and
to the absorption of molecular bands (in particular H$_2$O (high layers)
and CO).

\subsection{Ballistic model}

The formation of silicate dust is modelled for a fixed gas parcel from the outer layers of a pulsating star. We follow the parcel as it moves upward and downward and traces varying temperature and density conditions along its trajectory. We model the formation of dust within the parcel by calculating the nucleation rate by SiO and solving a set of moment equations for dust growth. From this we calculate the radiation pressure on the gas-dust mixture in the tight momentum coupling approximation.

In order to check the effect of radiation pressure on the dynamics of the flow we consider a ballistic model. It is based on the fact that in the envelope of a pulsating star the trajectory of a gas element follows for most of the time a ballistic trajectory; pressure effects become only important in the vicinity of the shock \citep{Ber85}. This allows one to start for a particular  pulsation phase $\Phi$ at some radius $r$ and take the velocity and density at that place and instant from the pulsation model. Then one solves the equation of motion for the gas parcel for these initial conditions to obtain the trajectory of the gas parcel. The local density in the parcel follows from the continuity equation in Lagrangian coordinates. The temperature at each step is determined by interpolation from the pulsation model. If the temperature has to be extrapolated beyond the model boundary we assume $T\propto r^ {-1/2}$ (see Sect. \ref{SectMiraMod}). {\rm The pressure follows from the ideal gas equation.}

The calculated trajectory closely follows the trajectory of that gas parcel in the pulsation model until the next shock overruns the gas parcel. Since shocks are not part of our simple ballistic model, we have to introduce them artificially by adding a suitable velocity jump $\Delta v>0$ to the velocity at the instant where the trajectory meets the next shock. The velocity increment $\Delta v$ required to follow the after-shock trajectory of the parcel in the pulsation model should ideally equal the velocity jump $\Delta v$ of the pulsation model. In practice we need a slightly higher value because the ballistic model neglects pressure effects in the vicinity of the shock and therefore does not completely reproduce the true motion in this limited region.

If radiation pressure is included in solving the equation of motion and if this force exceeds gravitational attraction from some point in time, the trajectory starts to level-off from a trajectory not including this force and then is rapidly driven out of the star. If this behaviour is detected we interpret this as an indication that an outflow is driven by dust condensation. It is, however, not possible to derive the properties of this outflow from our simplified model. We can presently only check if a pulsationally supported dust-driven outflow is principally possible.

The equations for the ballistic model are given in \citet{Fer06}. We consider for simplicity only the formation of a single amorphous silicate dust component. An iron-rich composition is assumed since presolar silicate grains from AGB stars have significant iron contents \citep{Vol09a,Ngu10,Bos10} and only rarely are iron-poor. The dust opacity required for calculating the radiation pressure is calculated with the complex index of refraction of the dirty silicate model of \citet[ their set 1]{Oss92}. The details of the opacity calculation are discussed in \citet{Gai13}. \citet{Bla12} argue that such iron bearing silicates cannot drive a wind because they vaporize at the close distances where the wind acceleration has to start. We note, however, that for silicates of lower iron content \citep[e.g. the $x=0.7$ data from ][]{Dor95} than used in their study ($x=0.5$) the radiative equilibrium temperature of the dust is lower by $\gtrsim150$ K such that this problem does not exist in our calculations.
 
\subsection{Nucleation by SiO}

The dust formation is treated as a two-step process: (1) Formation of seed particles from the gas phase and (2) precipitation of condensable material on the seed particles. 

Our basic assumption is that cluster formation by SiO molecules is  responsible for the onset of dust formation. We have discussed this in \citet{Gai13} and derived an empirically determined nucleation rate of
\begin{equation}
J_*=\left({p_{\rm g,SiO}\over kT}\right)^2\,\exp\left((1.33\pm3.1)-{(4.40\pm0.61)\times10^{12}\over T^3(\ln S)^2}\right)
\label{NukSioExp}
\end{equation}
(in units cm$^{-3}$\,s$^{-1}$) following the suggestions of \citet{Nut82}and using their experimental data, but now based on the re-determination of the vapour pressure of solid SiO by \citet{Wet12} and \citet{Gai13}, see also \citet{Nut06} and \citet{Fer08}. Here $S=p_{\rm vap}/p_{\rm g,SiO}$ is the supersaturation ratio, determined by the vapour pressure
\begin{equation}
p_{\rm vap}=\exp\left(-{49\,520\pm1\,400\,{\rm K}\over T}+32.52\pm0.97\right)
\label{VapPrSiO}
\end{equation}
of SiO (in cgs-units) and the partial pressure $p_{\rm g,SiO}$ of SiO in the gas phase. The latter has to be determined from a calculation of the composition of the gas-phase. Note that the formal similarity of Eq.~(\ref{NukSioExp}) with an equation from classical nucleation theory does not mean that we use classical nucleation theory. Our nucleation rate is a fit-formula to experimental data.

Figure \ref{FigNukPMod}b shows lines of constant values of the normalized nucleation rate $J_*/N_\mathrm{H}$ (with $N_\mathrm{H}$ being the density of H nuclei) in the $p$-$T$-plane and also the run of pressure and temperature in the outer layers of the Mira pulsation model for a number of different phases in the pulsation cycle. The circles show the experimental data to which the approximation for the nucleation rate is fitted. An application to astrophysical condensation conditions requires to apply the nucleation model, Eq.~(\ref{NukSioExp}), to a pressure regime with pressures about a factor of $10^{-4}$ times lower than the conditions under which the experimental data where acquired, while temperatures are about the same. Despite the uncertainties involved in such extrapolations we apply the empirical nucleation model since no better information is presently available.

The blue line corresponds to the onset of avalanche nucleation, defined here as that rate at which a few times 10$^ {-13}$ dust particles per H nucleus are formed within one year. This corresponds to the typical value of $7.6\times10^ {-13}$ dust particles per H-nucleus in outflows from oxygen-rich stars as derived by \citet{Kna85} from observations. For our numerical Mira pulsation model the outer layers of the model extend into the region where efficient nucleation of SiO is predicted by the empirical nucleation rate, Eq.~(\ref{NukSioExp}). That means that the trajectories of gas parcels moving up and down during the pulsation cycle traverse beyond $r\gtrsim 5R_\mathrm{p}$ a region of the $p$-$T$-plane where at the same time 
\begin{enumerate}
\item temperatures are low enough and 
\item densities are high enough
\end{enumerate}
for efficient SiO nucleation to be possible. 
 
\begin{figure*}

\resizebox{\hsize}{!}{\includegraphics[angle=-90]{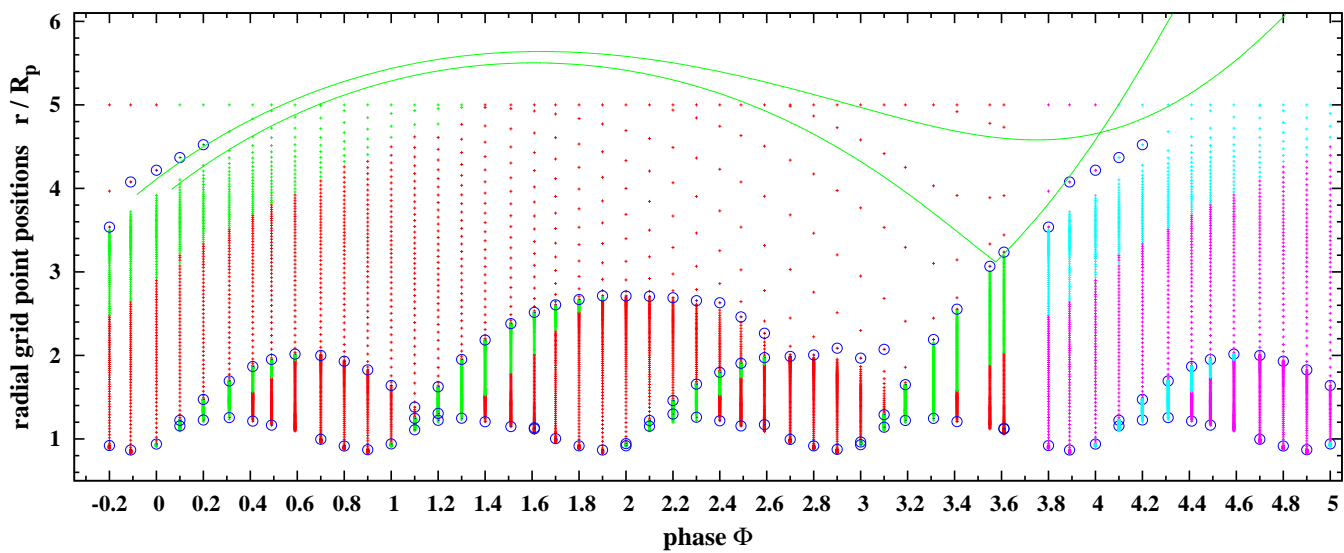}}

\caption{Trajectories of two different mass-element (solid green lines). The crosses correspond to the grid-points of the pulsation model \citep{Ire08, Ire11}. Green crosses correspond to outward directed velocity, red crosses to inward directed velocity. Circles roughly indicate the location of shock fronts. The kink in the lower trajectory is due to the shock hitting the gas parcel at phase $\Phi\approx3.55$.}

\label{FigTraj}

\end{figure*}

Unfortunately the outer part of the model calculation does not extend deep into the region beyond avalanche nucleation. This requires to consider for our calculation of dust condensation in a gas parcel trajectories that slightly extend beyond the outer border of the numerical pulsation model. The temperature is determined in that region by extrapolation assuming a $T\propto r^ {-1/2}$ dependency (see Sect. \ref{SectMiraMod}). Since no strong extrapolation is required, this is probably not critical.  

\begin{table}
\caption{Basic data used for calculating olivine condensation.}

\begin{tabular}{llrl}
\hline
\hline
\noalign{\smallskip}
Quantity &Symbol& Value & Unit \\
\noalign{\smallskip}
\hline
\noalign{\smallskip}
Molecular weight olivine & $A_{\rm d}$         & 172 & \\
Bulk density olivine     & $\rho_{\rm c}$      & 3.81 & g\,cm$^{-3}$ \\
Growth coefficient       & $\alpha$            & 0.1  & \\
Molecular weight SiO     & $A_{\rm SiO}$       & 44.09 & \\
Abundance of Si           & $\epsilon_{\rm Si}$ & $3.55\times10^{-5}$ & \\
\noalign{\smallskip}
\hline
\end{tabular}

\label{TabOliDat}
\end{table}

\subsection{Particle growth}

The second essential step for dust formation is the growth of seed particles to macroscopic dust grains. We presently do not know at which size the SiO clusters start to react with other species from the gas phase to incorporate additional oxygen to build the SiO$_4$-tetrahedron of silicates and to incorporate cations like Mg$^{2+}$ or Fe$^{2+}$ to form silicates. We assume here that this transition from cluster growth to silicate particle growth occurs already at rather small sizes such that the transition stage is not important for calculating the precipitation of the bulk of the silicate material.  

From laboratory investigations of silicate dust grains from AGB stars (found as  presolar grains in meteorites) we know that usually Mg-Fe-silicates are formed with widely varying iron content, where iron-poor grains are rare \citep{Vol09a,Ngu10,Bos10}. For simplicity it is assumed here that olivine with composition MgFeSiO$_4$ is formed. The corresponding basic chemical reaction is
\begin{equation}
\rm SiO\ +\ Mg\ +\ Fe\ +\ 3H_2O\ \longrightarrow\ MgFeSiO_4(s)\ +\ 3H_2\,.
\label{ChemEqCondOli}
\end{equation}
This is the net outcome of a series of reaction steps operating on the surface of a grain, the details of which are presently not known. The experimental observations with respect to the inverse process, evaporation of olivine, by \citet{Nag94} and \citet{Nag96} indicate that the attachment of SiO to the surface of a grain is the rate determining reaction step. The corresponding growth rate for a single dust grain then is (neglecting evaporation)
\begin{equation}
{\mathrm{d}\,a\over\mathrm{d}\,t}=V_1{\alpha p_{\rm g,SiO}\over
\sqrt{2\pi A_{\rm SiO}m_{\rm H}kT_{\rm g}}}
\label{EqPartGrow}
\end{equation}
where $a$ is the grain radius. Here  $\alpha_{\rm v}$ is the growth coefficient and $V_1=A_{\rm d}m_{\rm H}/\rho_{\rm c}$ the volume occupied by a chemical formula unit (atomic weight $A_\mathrm{d}$) in the solid condensed phase. The details are discussed, e.g., in \citet{Fer06} and \citet{GaS13}, the numerical values of the coefficients used for the model calculation are given in Table~\ref{TabOliDat}. The condensation coefficient $\alpha$ is still not accurately known. Experimental values between 0.12 and 0.025 are given, e.g. in \citet{Tac14}. We use here a value of 0.1 because lower values would result in too low dust formation efficiencies. 

The growth of an ensemble of dust grains and the radiation pressure on the dusty gas then are calculated by a moment method analogous to that described in \citet{Gai85}.

\section{Model results}

Figure \ref{FigTraj} shows, for the pulsation model, as small crosses the radial gridpoints of the model for a set of different phases $\Phi$ within the pulsation cycle.  The crosses are shown in green colour if the velocity at that point is directed outward and in red colour for inward directed velocity. The gridpoints where there is a high velocity jump to the next gridpoint are encircled. These points indicate the approximate position of a shock front running through the atmosphere.

From Fig.~\ref{FigNukPMod}b it is evident that efficient seed particle formation by SiO clustering requires that the temperature in the gas drops below about 800 K. From the run of temperature shown in Fig.~\ref{FigNukPMod}a it follows that during the upward and downward motion of a mass element in the pulsating envelope its trajectory has to reach its upper culmination point at a distance at least beyond four to five stellar radii to visit the region of sufficiently low temperature before falling back into higher temperature regions. 

A trajectory that accomplishes this requirement can be constructed for the pulsation model as follows: For some phase where the velocity is outward directed at the upper boundary at $5R_\mathrm{p}$ of the pulsation model, we solve the equation of motion for a gas parcel with the model velocity as initial condition, but in backward direction, until the temperature becomes too high for nucleation to be possible. Usually we take 0.8 of the initial radius as the endpoint of the inward integration. Then we start with the velocity obtained at this point as an initial condition an outward integration of the equation of motion simultaneously with the set of equations for dust formation and growth within the gas parcel. The integration is continued until one of the following cases is encountered:
\begin{enumerate}

\item The gas parcel is overrun by a shock front.

\item The gas parcel falls back to distances where all dust is evaporated.

\item Sufficient dust is formed such that radiation pressure drives the gas parcel out of the star.
   
\end{enumerate}
In the case where the ballistic trajectory crosses an outward running shock of the pulsation model, the velocity at the radius of intersection is changed by adding the velocity jump $\Delta v$ of the shock front to the instantaneous velocity. This case only happens if the velocity of the ballistic trajectory was inward directed before the shock arrived, and after addition of $\Delta v$ it is outward directed. Then a new outward integration is started.\footnote{%
In practice it turned out that a somewhat higher velocity increment $\Delta v$ is required than that determined from the velocity jump in the numerical model. This results probably from the fact, that the shock is numerically broadened and its precise position is difficult to determine from the numerical solution. The velocity jump in the numerical model then usually underestimates the true velocity jump at the true shock position.} 
This is continued until one encounters one of the three cases and so on.

\begin{figure}

\resizebox{.97\hsize}{!}{\includegraphics[angle=-90]{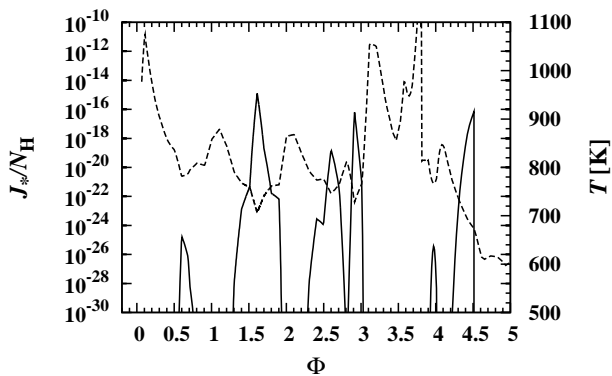}}

\caption{Variation of gas temperature (dashed line) and dust particle formation rate (solid line) in the gas parcel along the lower trajectory in Fig.~\ref{FigTraj}. The abscissa is the phase of the pulsation. The left ordinate is the dust particle formation rate in particles per H nucleus and per second. }

\label{FigTJ}
\end{figure}

Figure \ref{FigTraj} shows two such trajectories. The gas parcel crossing the boundary $r=5R_\mathrm{p}$ at phase $\phi=0.49$ in the model forms sufficient dust to be expelled by radiation pressure already before the next shock arrives. The parcel crossing the boundary at phase $\phi=0.70$ falls back until it is hit by a shock.  The starting point in both cases was searched for by integrating backwards in time from $r=5R_\mathrm{p}$ to $r~\sim4R_\mathrm{p}$ and then all equations inclusive dust formation are solved forward in time. Note that the fact that the two trajectories in Fig. \ref{FigTraj} do intersect results from the fact that our simplified model does not account for the dynamical back-reaction of the dust on hydrodynamics. In a consistent model such a situation would either be avoided or a shock would be created.

By continued dust growth, radiation pressure from some instant exceeds the gravitational pull by the star and the gas parcel finally is expelled. The wind-driving capability of the dust can only be seen in the accelerated outward motion of dust relative to the gas because acceleration of the gas by frictional coupling to dust is not included in the ballistic model.

Figure \ref{FigTJ} shows for the lower trajectory in Fig.~\ref{FigTraj} the variation of the gas temperature during the course of the motion of the gas parcel. The temperature appears strongly fluctuating with phase. This is a property of the temperature in the outer layers of the pulsation model, which is also apparent in Fig.~\ref{FigNukPMod}. The figure also shows the variation of the normalized seed particle formation rate $J_*/N_\mathrm{H}$ during the course of the motion of the gas parcel. The rise and fall of the nucleation rate seen in the figure reflects the temporary excursion into the low temperature range. 

Around maximum outward excursion of the gas element the temperature becomes rather low and there is a burst of seed particle  formation. Additional but less strong bursts occur at some later low-temperature episodes. Overall, seed particles are formed in this model intermittently. Beyond maximum outward excursion the temperature raises again during fall-back of the parcel. It does not increase, however, to such high values that the freshly formed dust disappeares by evaporation before the next shock arrives; this would require temperatures above $\approx 1\,100$ K. Instead the next shock throws the element back to low temperatures. As Fig.~\ref{FigTJ} shows there starts a new period of nucleation, which is inconsequential, however, because the bigger grains formed during the first nucleation period preferentially take up the condensable material.

\begin{figure}

\resizebox{.9\hsize}{!}{\includegraphics[angle=-90]{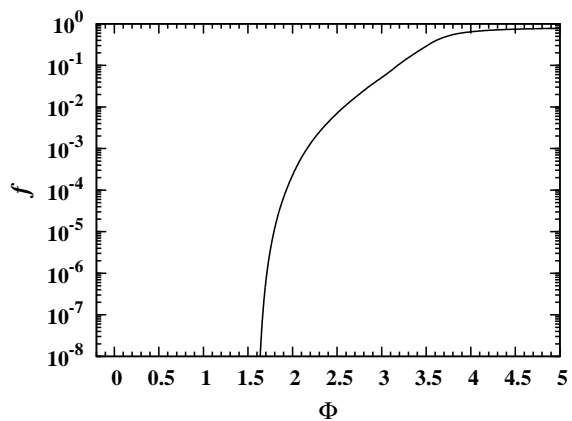}}

\caption{Dust condensation along the lower particle trajectory in Fig.~\ref{FigTraj}. The abscissa is the phase of the pulsation. The ordinate is the fraction of the Si condensed into silicate dust. }

\label{FigFcond}
\end{figure}

Figure \ref{FigFcond} shows the variation of the fraction $f$ of all Si that is condensed into silicate grains (the degree of condensation) during the course of the motion of the parcel. After the first burst of seed particle formation the degree of condensation $f$ rapidly increases but remains rather low because in a stellar envelope the particle densities are low if the temperature is low, such that the rate at which condensable material precipitates on the seed particles remains low. It is only if the gas parcel falls back and the gas is compressed and particle densities increase again that the growth rate progressively increases. This results from the fact that, as long as one is not close to the evaporation limit of the bulk condensate, particle growth depends more strongly on density than on temperature, such that re-heating after seed particle formation only suppresses nucleation, but not particle growth. The final levelling off of the condensation curve seen in the figure results from the rapid expansion of the gas after shock arrival.

\section{Discussion}
\label{SectConclu}

The calculation shows that, even though seed particle formation by SiO requires rather low gas temperatures ($\lesssim800$\,K), this process results --- if applied to a pulsation model with properties like $o$ Ceti --- in sufficient final dust production to drive a gas parcel out of the star. This is the necessary pre-requisite in order that nucleation by the abundant SiO molecules and subsequent growth of magnesium-iron silicates on these seed particles could be responsible for pulsation-supported dust-driven winds for Miras.
 
\begin{table*}

\caption{Distance, $R$, in units of the photospheric radius, $R_\mathrm{p}$, and temperature of inner edge, $T_\mathrm{i}$, of silicate and aluminium dust shell as derived from interferometer observations of Mira variables.}

\begin{tabular}{l@{\hspace{1cm}}ccccccccc}
\hline
\hline
\noalign{\smallskip}
Star & \multicolumn{2}{c}{$o$ Cet\tablefootmark{a}} & VX Sgr\tablefootmark{a} & IK Tau\tablefootmark{a} & RR Aql\tablefootmark{b} & S Ori\tablefootmark{b} & R Cnc\tablefootmark{b} & \multicolumn{2}{c}{GX Mon\tablefootmark{b}} \\
dust type & silicate & silicate & silicate & silicate & silicate & aluminium & aluminium & aluminium & silicate \\
\noalign{\smallskip}
Phase   & Min. &  Max & Min & Max & \\
\noalign{\smallskip}
\hline
\noalign{\smallskip}
$R_\mathrm{i}/R_\mathrm{p}$ & 3.0   & 3.0  & 4.6 & 5.5 &  4.6 &  1.9 &  2.2 & 2.1 & 4.6 \\
$T_\mathrm{i}$ & 1060 & 1280 & 720 & 990 & 1130 & 1340 & 1210 & 1350 & 1070\\
\noalign{\smallskip}
\hline
\end{tabular}
\tablefoot{
\tablefoottext{a}{\citet{Dan94}},
\tablefoottext{b}{\citet{Kar13}}
}

\label{TabRiTi}
\end{table*}

In this model a lot of dust particles are found at distances closer to the star than the distance of the layer where the seed particles for dust are formed.  The initial growth from seeds to macroscopic grains occurs on the descending part of the trajectory of a gas element that executes the up and down motions in the outermost layers of a pulsating star. The downward motion moves the dust closer to the star where it becomes hotter and may evaporate if it is not blown out as in the cases shown in Fig.~\ref{FigTraj}. The hottest dust found in the observed infrared spectrum of the star is then not related to the dust temperature in the dust forming layer. It is determined either (1) by the smallest distance from the star to which a gas parcel falls back before it is driven out by radiation pressure, or (2) by evaporation of the freshly formed dust. In view of the strongly fluctuating temperature conditions at several stellar radii (see Fig.~\ref{FigNukPMod}) there is no well defined inner edge of the dust shell but only a transition zone with variable dust content. 

For the model considered here this zone extends from about~3 to about roughly 6 stellar radii. It would then depend on the epoch of observation at which layer one would find the inner edge of the silicate dust shell. Additionally one would expect from the highly variable temperature structure of the outermost layers (cf. Fig.~\ref{FigNukPMod}) and, in fact, one observes \citep[e.g.][]{Lop97,Wei06} some clumpiness of the dust resulting in a variable dust content at $R/R_\mathrm{p}\gtrsim3$. Observationally it is found that temperature and location of the inner edge of dust shells around M giants vary in a wide range (see Table \ref{TabRiTi}). In this respect the processes of SiO-nucleation as  trigger of silicate dust formation is compatible with what one observes, despite the somewhat low temperature where avalanche nucleation starts.   

One remark is appropriate at this point. The concept of dust condensation temperature as it is generally used by observers corresponds to hottes dust temperatures required to explain the observed spectrum; this is the temperature given in Table \ref{TabRiTi}. This is not identical with the physical concept of dust formation temperature which means the gas temperature at which avalance nuleation commences. In a stationary outflow both temperatures should be similar, except for possibly differing greenhouse effects of dust and gas, but for pulsating stars there may be considerable differences between the two temperatures as we have just argued. The fact that in our model dust forms at lower temperatures than the observed hottest dust temperature results from the different meaning of both temperatures. The precise relation between the two temperature concepts can only be derived by detailed consistent models, which are out of our scope.

Another important aspect of our model is that the onset of wind driving by radiation pressure on dust will occur in a zone closer to the star than the zone where dust nucleation is active. In the discussions on the problem of  wind-driving by iron-bearing silicate dust \citep{Bla12} it is assumed that these zones essentially coincide which may result in the problem that iron-bearing dust is too hot.

An alternative mechanism for seed particle formation for silicate dust growth is the formation of small particulates from highly refractory Ca-Al-compounds which serve as growth centres for silicate dust. Such aluminium compounds are expected to nucleate and grow at significantly higher temperature and much closer to the star than silicates. The precipitation of a new solid phase on an already existing surface requires no substantial supersaturation and usually commences close to the stability limit, which is about 1\,100\,K for silicates under circumstellar conditions \citep{GaS13}. 

Observationally it has been found that in some objects there seems to exist some dust already at distances of about two stellar radii \citep[see][ and references therein]{Kar13,Kar15} which is interpreted as to be due to corundum dust grains. Table \ref{TabRiTi} shows for a few oxygen-rich Miras where such data have been reported the inner radii and dust temperatures of spatially resolved dust layers close to the star as derived from interferometer studies of the emission from the distance range $R\lesssim10 R_\mathrm{p}$. Observationally, thus, one finds objects which seem to form silicate dust only, aluminium dust only, or both species. This shows that two different dust nucleation processes may operate in such stars, but the special circumstances under which one or the other of these is active have still to be explained. The few data for silicate dust temperatures at the closest observed distance to the star seem to indicate that both, silicate growth on top of aluminium dust and separate silicate dust nucleation occurs in Miras. In the latter case the data seem to be compatible with our findings on SiO nucleation.
 
The reaction knetics of dust formation was discussed recently by \citet{Gou12} and by \citet{Gob16}. This approach suffers presently from the problem that only insufficient information is available on cluster properties and reaction rates. These model are substantially based on MgSiO-clusters which, however, are not found in mass spectroscopic investigations of the vapour of forsterite \citep{Nic95}. The model of \citet{Gob16} additionally includes (SiO)$_n$ clusters up to $n=4$ so that the SiO-clustering considered in the present paper is to some extent also included in their model. A similar kinetic model using (SiO)$_n$ clusters up to $n=10$ \citep[ Sect. 10.5.4.3]{GaS13} using a similar set of rate coefficients showed, however, that the resulting condensation temperature is much lower than suggested by the experimental results of \citet{Nut82}. Hence the kinetic data are presently not yet sufficiently accurate to calculate nucleation rates by this approach.

The results of this preliminary study demonstrate that, using experimental results of \citet{Nut82}, SiO nucleation  explains observed properties of dust formation at least for some of the oxygen-rich Miras. A self-consistent implementation of the process into pulsation models is required, however, to elucidate the interplay between nucleation of silicate and aluminium dust and to obtain results that can be compared more directly with observations.


\begin{acknowledgements}
This work was supported in part by `Schwerpunktprogramm 1385' supported by the `Deutsche Forschungs\-gemeinschaft (DFG)'.
\end{acknowledgements}


\end{document}